\documentclass[aps,prl,superscriptaddress,twocolumn,longbibliography]{revtex4-1}
\usepackage[pdftex]{graphicx}
\usepackage{color}
\usepackage{footmisc}
\usepackage{amsmath}
\usepackage{subfigure}

\usepackage{setspace}

\usepackage{amsmath}
\usepackage{xfrac}
\usepackage{xcolor}

\usepackage{dcolumn}   
\usepackage{amsfonts}
\usepackage{booktabs}
\usepackage{siunitx}
\usepackage{multirow}
\usepackage{mathtools}

\graphicspath{{./figures/}} 
\graphicspath{{./pictures/}}

\newcommand{\UOT}{Department of Materials Science and Engineering, University of Toronto, Toronto, ON, Canada}
\newcommand{\QUEENS}{Department of Mechanical and Materials Engineering, Queen's University, Kingston, Ontario, Canada}

\begin{document}

\title{The effect of He on the order-disorder transition in Ni$_3$Al under irradiation}



\author{Peyman Saidi}
\email[]{peyman.saidi@queensu.ca}
\affiliation{\QUEENS}

\author{Pooyan Changizian}
\affiliation{\QUEENS}

\author{Eric Nicholson}
\affiliation{\UOT}

\author{He Ken Zhang}
\affiliation{\QUEENS}

\author{Yu Luo}
\affiliation{\QUEENS}

\author{Zhongwen Yao}
\email[]{yaoz@queensu.ca}
\affiliation{\QUEENS}

\author{Chandra Veer Singh}
\affiliation{\UOT}

\author{Mark R. Daymond}
\email[]{mark.daymond@queensu.ca}
\affiliation{\QUEENS}

\author{Laurent Karim B\'{e}land}
\affiliation{\QUEENS}

\date{\today}

\begin{abstract}

The order-disorder transition in Ni-Al alloys under irradiation represents an interplay between various re-ordering processes and disordering due to thermal spikes generated by incident high energy particles. Typically, ordering in enabled by diffusion of thermally-generated vacancies, and can only take place at temperatures where they are mobile and in sufficiently high concentration. Here, \textit{in-situ} transmission electron micrographs reveal that the presence of He--usually considered to be a deleterious immiscible atom in this material--promotes reordering in Ni$_3$Al at temperatures where vacancies are not effective ordering agents. A rate-theory model is presented, that quantitatively explains this behavior, based on parameters extracted from atomistic simulations. These calculations show that the $\mathrm{V_2He}$ complex is an effective agent through its high stability and mobility. It is surmised that immiscible atoms may stabilize reordering agents in other materials undergoing driven processes, and preserve ordered phases at temperature where the driven processes would otherwise lead to disorder.

\end{abstract}


\maketitle



In physical metallurgy, ordered phase precipitation is commonly employed as a strengthening strategy. This is one of many examples of an ordered phase providing technologically relevant properties to a material \cite{cahn1994place}. Driven processes--including neutron irradiation and extreme deformation--act as disordering agents. They inject energy into the system, which allows it to escape equilibrium conditions. This driving force towards disorder is counteracted upon by micro-mechanisms that favor the lower-energy ordered phases. In metals and alloys, these mechanisms are typically associated to vacancy diffusion.

A figure case of such an interplay is the disordering of $\gamma '$ Ni$_3$Al phases in X-750--a Ni-based superalloy \cite{ewert1998ion}--under neutron irradiation. From a practical point of view, this interplay is critical to the long-term performance and safety of CANDU reactors, where this alloy is used for garter springs separating pressure tubes from calandria tubes. In this reactor concept, the garter springs are situated less than 1 cm away from the closest fuel pellet, and therefore undergo a significant amount of neutron irradiation (typically $\sim$ 4 dpa/year). Notably, the high flux of thermal neutrons?which confers exceptional neutron economy to CANDU reactors-breeds significant concentrations of He, $\sim$1500 appm/year, as the neutrons are absorbed by Ni$^{59}$ \cite{griffiths2014effect}.

The competition between radiation-induced disordering and vacancy-induced ordering of $\gamma '$ has been extensively studied \cite{potter1976irradiation,bourdeau1994disordering,zhang2013microstructural,zhang2014stability} and predictive models now explain the set of conditions under which the precipitates remain ordered and those under which they do not \cite{martin1984phase, matsumura1996formation, abromeit1999modelling}. Molecular dynamics and ion irradiation studies indicate that disordering is caused by thermal spikes induced by collision cascades  \cite{spaczer1994computer, spaczer1995evidence, gao2000temperature, ye2010atomistic, zhang2014radiation, lee2015atomistic}. Rate theory, on-the-fly kinetic Monte Carlo and ion irradiation experiments indicate that re-ordering is caused by mono-vacancies \cite{ewert1998ion, oramus2003ordering, martinez2016atomistic}. Reordering can only take place at temperatures where mono-vacancies are sufficiently mobile, typically above 750 K. The effect of He on the order-disorder competition has--to the best of our knowledge--not been studied.

In this Letter, we present experimental and theoretical work that explores the effect of He on the order-disorder transition in Ni$_3$Al $\gamma '$. Ion-irradiated samples--used as surrogates for neutron-irradiated samples--were characterized by transmission electron microscopy (TEM). On the theoretical side, atomistic simulations, including electronic density functional theory (DFT), molecular dynamics (MD), the Activation Relaxation Technique \emph{nouveau} (ARTn) and the nudged elastic band (NEB) simulation are used to explore the energetics and kinetics of the Ni$_3$Al ordering micro-mechanisms in the presence of He. These quantitative results are used as inputs to a rate-theory model to predict order-disorder transitions in Ni$_3$Al under different temperature and irradiation conditions.

\begin{figure}
	\centering
	\includegraphics[width= 0.5\textwidth]{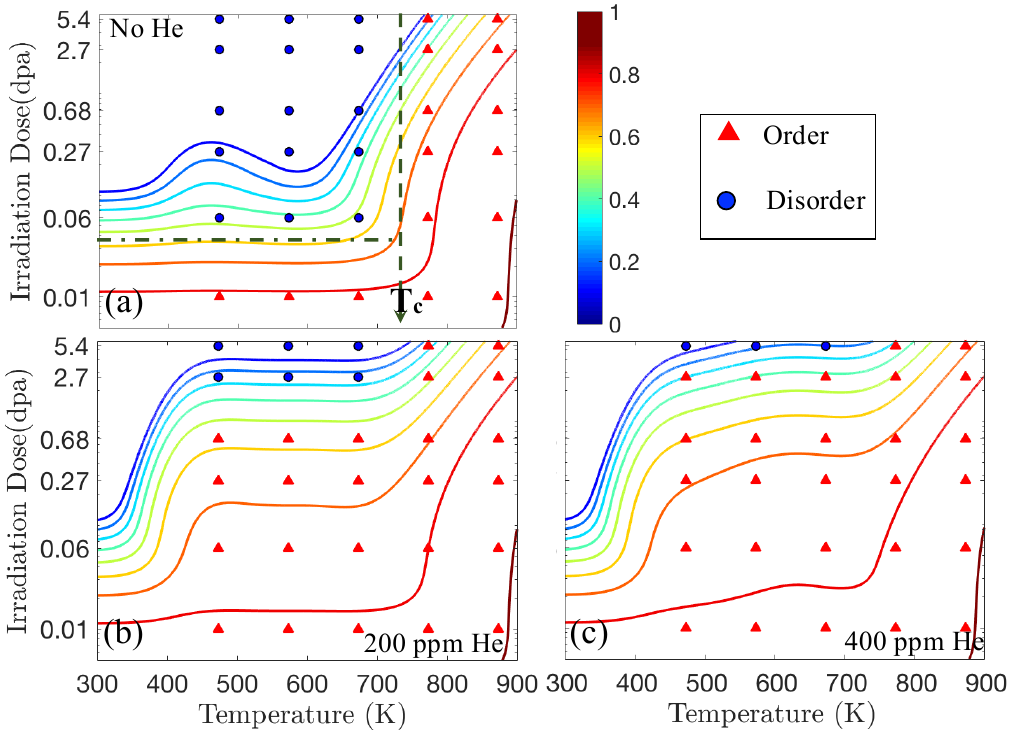}
	\caption{\small The order parameter of the $\mathrm{Ni_3Al}$ precipitates, as a function of He concentration, irradiation dose and temperature. Red triangles and blue circles indicate the experimental presence of ordered and disordered structures, respectively. The lines are the predictions of a rate-theory model, calibrated using atomistic simulations. Results at three He concentrations are reported: (a) zero, (b) 200ppm and (c) 400 ppm.  $\mathrm{T_c}\sim 740$ K, corresponding to the dashed line in the "0 ppm He" graph, beyond which disordering was not observed.} 
	\label{Fig:Experiment}
\end{figure}

The experimental setup is as follows. The three sample sets were pre-implanted at room temperature with He concentrations of 0, 200 and 400 ppm. \emph{In situ} TEM heavy-ion (1 MeV Kr$^{+2}$) irradiation was performed up to doses of 5.4 displacements per atom (dpa) at five different temperature points ranging from 473 to 873 K. The presence of the [001] superlattice reflection peak was used to measure the presence of the $\gamma '$ phase. In the absence of such a peak, the Ni$_3$Al was considered as being disordered.  Fig. \ref{Fig:Experiment} reports these measurements, alongside our model's predictions. The model and its physical origins are described in the following pages. The red triangles represent the ordered structure--superlattice reflections observed, whereas the blue circles represent the disordered structure--no superlattice reflection observed. More experimental details are available in the Supplementary Information (SI).

The experiments show that He pre-implantation promotes ordering of the $\gamma '$ phase. Without He, at temperatures lower than T$_c$=740 K, an irradiation dose between 0.01 and 0.06 dpa is sufficient to eliminate order in the $\gamma '$phase. Pre-implanting with He significantly raises the dose threshold required to induce disorder. Adding 200 (400) ppm of He elevates the dose threshold to a level between 0.68 (2.7) and 2.7 (5.4) dpa (respectively).

Consistent with previous reports \cite{ewert2000reordering}, at temperatures hotter than T$_c$= 740 K, order parameter of at least 0.3 is preserved at all irradiation doses tested. Beyond this critical temperature, the mono-vacancies are mobile and promote order. Our experiments reveal that He stabilizes the ordered phase at lower temperatures, where mono-vacancies are not sufficiently mobile, nor in great enough concentration, to do so. This suggests that He stabilizes another ordering agent, which is more mobile than mono-vacancies.

\begin{table}[]

	\caption{\small Formation, binding and migration energies of various point and cluster defects in $\mathrm{Ni_3Al}$ calculated via DFT. The migration energies for the He-V cluster via both the direct exchange and ring exchange mechanisms are shown.}
\begin{tabular}{ccc}
\hline
Configuration & $E_f$(eV)                     &$E_m$(eV)                                          \\ \hline
$\mathrm{V^{Al}\xrightarrow{\text{Al}} V^{Ni}}$ 		           &1.64  &0.99  \\
$\mathrm{V^{Ni}\xrightarrow{\text{Ni}} V^{Al}}$		           &3.04 &0.96 \\ \hline
\multicolumn{1}{l}{}           & \multicolumn{1}{l}{$E_b$(eV)} & \multicolumn{1}{l}{}  \\ \hline
$\mathrm{V_2^{Dis}(1nn)\rightarrow V_2^{Ord}(1nn)}$	   &0.29 &0.51 \\ \hline
$\mathrm{V_2He^{Dis}(1nn)\rightarrow V_2He^{Ord}(1nn)}$&0.51 &0.62  \\
$\mathrm{V_2He^{Dis}(1nn)\rightarrow V_2He^{Ord}(2nn)}$&0.18 &0.62 \\
$\mathrm{V_2He^{Dis}(1nn)\rightarrow V_2He^{Ord}(3nn)}$&0.04 &0.52 \\
$\mathrm{V_2He^{Dis}(2nn)\rightarrow V_2He^{Ord}(1nn)}$&0.51 &0.84  \\
$\mathrm{V_2He^{Dis}(3nn)\rightarrow V_2He^{Ord}(1nn)}$&0.51&0.59 \\ \hline
\end{tabular}
\label{EM}
\end{table}

In FCC materials, di-vacancies are generally more mobile than mono-vacancies, which can be explained by bond-breaking arguments \cite{johnson1966calculations}. Also, their binding energy is typically much lower than their migration energy, which means that they tend to dissociate rather than migrate. \textit{Ab initio} calculations confirm that Ni$_3$Al behaves in this way. Formation and binding energies of point defects and small clusters assessed using DFT calculations, performed with the Vienna \emph{ab initio} simulation package (VASP)  \cite{kresse1993ab, DFTSETTING} \nocite{perdew1996generalized,kresse1999ultrasoft,monkhorst1976special}. The simulation details are provided in the SI. The results are presented in Tab. \ref{EM}. The binding energy of the di-vacancies in first nearest-neighbor (1NN) position (0.29 eV) is smaller than the migration barrier of the di-vacancy from a disordered to ordered configuration (0.51 eV). The addition of He stabilizes the di-vacancy. The V$_2$He complex with vacancies in 1NN position has a binding energy of 0.51eV. 

In order for this V$_2$He to be an effective reordering agent, its migration energy must be sufficiently small. The mobility of the V$_2$He complexes was calculated using DFT-based NEB calculations. In order to generate trial minimum energy paths, ARTn \cite{mousseau1998traveling,malek2000dynamics,machado2011optimized,mousseau2012activation,trochet2019off, ARTn}\nocite{barkema1996event,munro1999defect,henkelman1999dimer,cances2009some,marinica2011energy,n2015probing,dufresne2018atomistic,beland2015interstitial,osetsky2018existence} searches were performed. The inter-metallic interactions are based on Ref.\cite{skirlo2012role}. The He-Ni and He-He interactions are based Ref. \cite{torres2017atomistic}. The He-Al interactions were calculated by combining ZBL universal potential for short distances and DFT calculations \cite{EAMPot}\nocite{mishin2004atomistic,beck1968new,zhang2016development,Giannozzi_2009,ziegler1985stopping}. Initial states (V$_2$He complexes in a disordered Ni$_3$Al simulation box) are provided to ARTn, which generates a set of transition states and final states. The ARTn-generated transitions with the smallest activation energies were used as inputs for DFT-based NEB calculations. For comparison, the activation barriers of mono-vacancy jumps are presented as well. Computational details can be found in the SI.

The results of the NEB calculations are reported in Tab. \ref{EM}. The mono-vacancy migration barrier is greater than 0.96 eV, in agreement with the literature \cite{gopal2012first}. This corresponds to an antisite Ni diffusing to 1NN and leaving a vacancy on the Al sublattice. Assuming a standard 10 THz prefactor, the waiting time at 600 K is approximately 1$\mathrm{\mu s^{-1}}$.

\begin{figure}
	\centering
	\includegraphics[width= 0.5\textwidth]{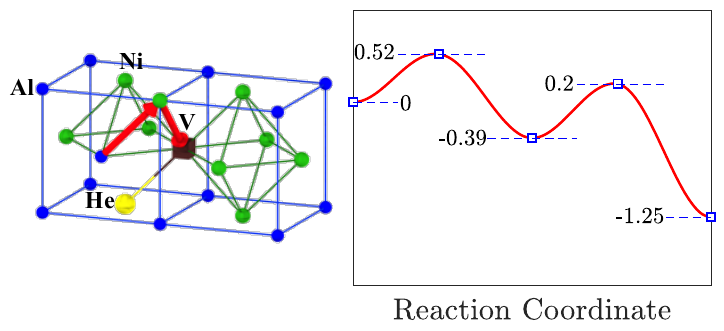}
	\caption{\small Left panel: a reaction path from the disordered sublattice to the ordered sublattice, involving a V$_2$H complex. The formation and migration energies of the complex are reported in Table \ref{EM}, among other plausible paths. The $\mathrm{V_2He(1nn)\xrightarrow{\text{Ord.}} V_2He(3nn)\xrightarrow{\text{Ord.}}  V_2He(1nn)}$ mechanism is illustrated here. The displacement vector is colored red. The moves result in ordering without changing the configuration of the $\mathrm{V_2He}$ complex. He and V partially dissociate to the 3nn in the first jump and rebond in the second jump. Right panel: a schematic representation of the minimum energy path along this two-step reaction path.} 
	\label{Fig:OrderingGraph}
\end{figure}

Our ARTn-inspired DFT-based NEB calculations revealed a set of jump sequences involving the V$_2$He which leads to ordering of the $\gamma '$ phase, without changing its configuration. One such move is illustrated in Fig. \ref{Fig:OrderingGraph}. A schematic representation of the potential energy landscape along the reaction path is also provided. In this move, the He and V partially dissociate to the 3rd nearest neighbor position (3nn) in the first jump, and rebond in the second jump. The process results in ordering, leading to a 1.25 eV decrease of the energy. The highest activation energy barrier crossed during the two-jump sequence is 0.59 eV. In first approximation, this is corresponds to the activation energy of the V$_2$He-enabled reordering process--note that it is of the same order of magnitude as the complexes' binding energy. The waiting time at 600 K is of the order of 10$\mathrm{n s^{-1}}$. This rapid jump rate, much faster than that of mono-vacancy jumps, could explain why He promotes order at temperatures lower than T$_C$.

The main takeaway from the DFT calculations is that V$_2$He complexes have binding energies and mobilities that make them a plausible ordering agent. However, in order to build a predictive model, the probability of occurrence of ordering due to mobility of each agent, as well as the agents concentration, must be determined. The assumptions and equations used to estimate this concentration are explained in the following paragraphs. 

V-He complexes tend to cluster into bubbles. New complexes are released by collision cascades. Afterwards, their concentrations change due to: clustering, sinking, pinning by He atoms and dissociation. At time $\tau$, the concentration of species $s$ is $C_s=C_{0}f_{s}-\sum_{i=1}^{m}\left.\int_0^\tau \frac{\mathrm{d}C_{s}}{\mathrm{d}t}\right|_{i}$, where $C_{0}$ is the total concentration of generated defects or super saturated concentration, $f_{s}$ is the fraction of the species of interest, and $\left.\int_0^\tau \frac{\mathrm{d}C_{s}}{\mathrm{d}t}\right|_{i}$ represents the change of concentration of species $s$ due to mechanism $i$. Let $\lambda_{d-s}$ be the distance that a defect travels before being trapped at a sink. A mean diffusion time $t=\lambda_{d-s}^2/D_d$ is needed for the vacancy to sink at the trap, where $D_d$ is its diffusivity. Therefore, for a given defect $d$, the variation of the concentration due to clustering is $\left.\frac{\mathrm{d}C_{s}}{\mathrm{d}t}\right|_{clustering}=-\frac{C_{s}D_{d}}{\lambda_{d-s}^2}$. Similarly to the logic of clustering, pinning occurs by diffusion of He. A mean diffusion time $t=\lambda_{He-d}^2/D_{He}$ is needed for vacancy trapping, where $D_{He}$ is the He interstitial diffusion coefficient. It should be noted that the diffusion rate of V and $\mathrm{V_2}$ are negligible compared to that of He atoms. Finally, according to the first order dissociation model, the dissociation rate of a He-V complex is expressed as \cite{morishita2003thermal}
$\left.\frac{\mathrm{d}C_{s}}{\mathrm{d}t}\right|_{dissociation}=-C_s\nu_0\exp\left(-\frac{E_b+E_m}{k_bT}\right)$, where $\nu_0$ is the attempt frequency, $E_b+E_m$ is the energy for dissociation.

MD simulations of collision cascades in the presence of He were performed to provide an estimate of irradiation-induced vacancy cluster and V-He complex distributions. Also, we observed that the number of irradiation-induced vacancies depends on the concentration of He and that all interstitial He atoms are pushed to substitutional sites by collision cascades. 60\%-75\% of the vacancies are in the form of mono-vacancies and 15\% of them are $\mathrm{V_2He}$ (Fig SI.4). More information about the collision cascades can be found in the SI \cite{CASCADE}\nocite{beland2016atomistic,stoller2016impact,beland2017accurate,trung2018threshold,li2012defect,adams1989formation,morishita2006mechanism}.

\begin{figure}
	\centering
	\includegraphics[width= 0.5\textwidth]{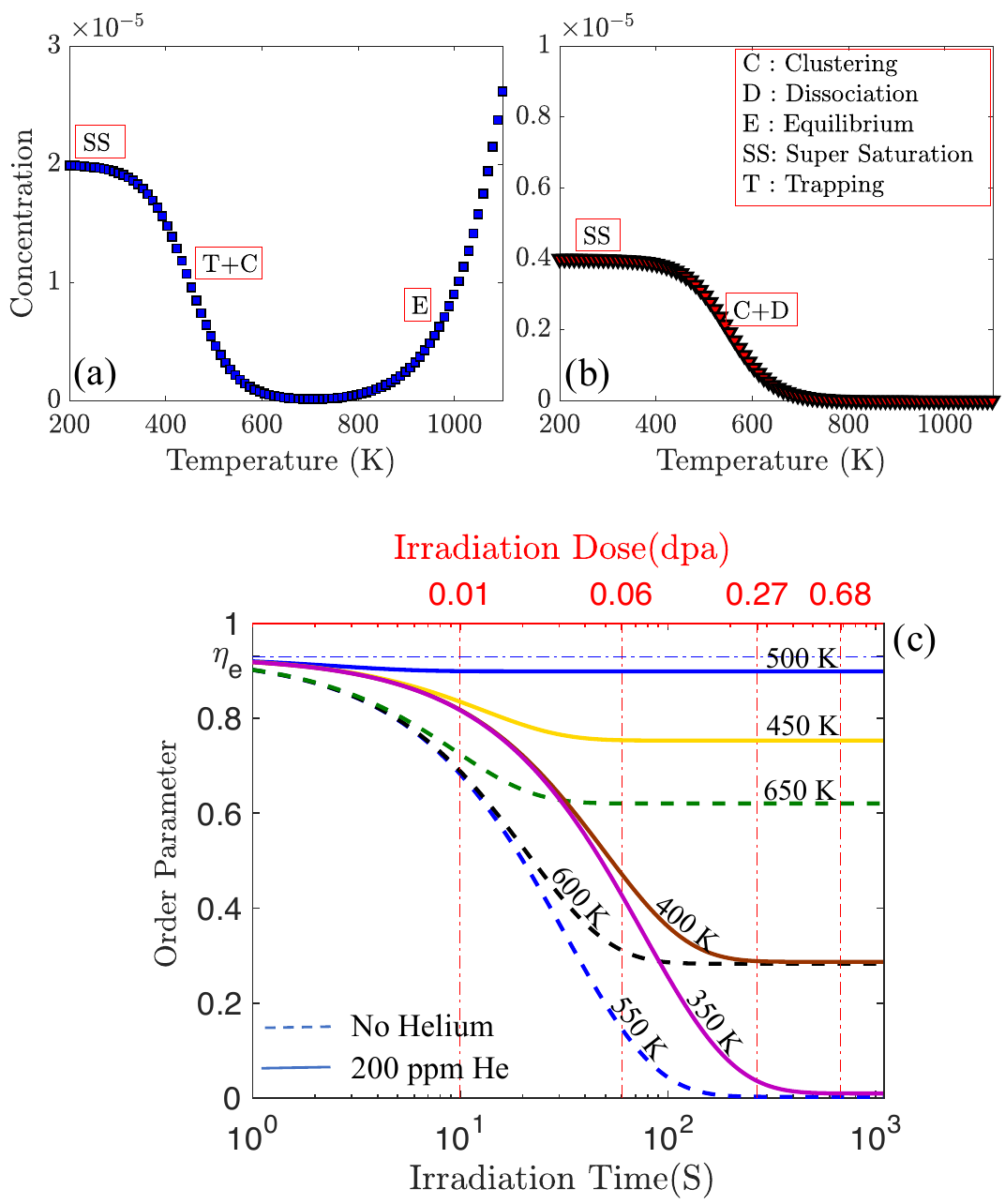}
	\caption{\small (a) The temperature-dependent, steady-state mono-vacancy concentration predicted by the model. (b) The temperature-dependent, steady-state $\mathrm{V_2He}$ concentration predicted by the model.(c) The time (or, equivalently, dose) dependence of the $\mathrm{Ni_3Al}$ precipitate order parameter, as predicted by the model. The values are reported at 350K, 400K, 450K, and  500K in the system without He and with 200 appm He.} 
	\label{Fig:CandEta}
\end{figure}

Using the equations above and the MD-informed irradiation-induced vacancy and V-He complex production rates, the steady state concentration--when the defect generation rate is equal to removal rate--is calculated at each temperature. The details are provided in the SI. The temperature dependence of the steady-state concentrations of mono-vacancies and V$_2$He are plotted in Fig. \ref{Fig:CandEta} (a) and (b), respectively. The concentration of mobile mono-vacancies is mostly thermally-controlled, the concentration of mobile V$_2$He complexes is chemically controlled (i.e. controlled by the concentration of immiscible He).

Liou and Wilkes \cite{liou1979radiation} proposed a formulation to calculate the order parameter $(\eta)$ as a function of irradiation dose and temperature in the absence of He. Here, it is extended to include the effect of He.  The disorder-order transition is considered as a reversible chemical reaction due to an interplay between radiation-induced disordering and thermal/chemical reordering. The rate of disordering is simply a function of disordering efficiency and irradiation dose \cite{lee2015atomistic}. According to Nowick et. al \cite{nowick1958simple}, the thermo-chemical reordernig  is a function of the equilibrium order parameter and the re-ordering rate constant defined as:
 \begin{equation}\label{Eq:rateConstant}
k_0=\left(\frac{X_{Ni}}{X_{Al}}\right)^{0.5}Z_{\beta}\exp\left(-\frac{V_0}{2k_bT}\right)\sum_{j=1}^{n}\exp\left(-\frac{U_j}{k_bT}\right)
\end{equation}

where $X_i$ is the atomic fraction of component $i$, $V_0$ is the activation energy when $\eta=1$ and $Z_{\beta}$ is the number of $\beta$ sites (corners of FCC unit cell) which are the nearest neighbours to the $\alpha$ site (the centre of FCC unit cell faces). The term $\exp\left(-U_j/k_bT\right)$ is the Boltzmann factor representing the probability of occurrence of mechanism $j$ with activation energy $U_j$. Since more than one ordering mechanisms are active simultaneously, the overall transition is the summation of all effective mechanisms. The Boltzmann factor for each reaction is limited to a unique atomic configuration. For instance  $\mathrm{V_2He^{Dis}(1nn)\rightarrow V_2He^{Ord}(1nn)}$ is active if both vacancies of a di-vacancy are in the nearest neighbour of the atom in the wrong sublattice. This restriction limits the choice of configurations to $\frac{2}{12}Z_{\alpha}$ and $\frac{2}{12}Z_{\beta}$ for Ni and Al in the wrong sublattices, respectively. Therefore, the Boltzmann factor for the $\mathrm{V_2He}$ is: $C_{V_2He}\exp\left(-\frac{E_m^{V_2He}}{k_bT}\right)\left(X_{Ni}\frac{2Z_{\alpha}}{12}+X_{Al}\frac{2Z_{\beta}}{12}\right)$

In this equation, the essential terms are migration energy, calculated using DFT, and the concentration of the agent of interest explained above. For sake of completeness, di-vacancies and self-interstitial atoms were also included. The contribution of the latter two to reordering are negligible, because di-vacancy either rapidly dissociate unless stabilized by He, and self-interstitial-vacancy recombination can cause only a single re-ordering event, while V$_2$He and V jumps lead to multiple re-ordering events. The expressions that define their behavior are given in the SI. Order parameter is plotted as a function of irradiation irradiation time, according to the model, in Fig. \ref{Fig:CandEta} (c), at different temperature, with and without He. The steady state order parameter increases in the presence of He.

The model and experiment plotted Fig. \ref{Fig:Experiment} are in close agreement. Note that the critical irradiation dose $ID_c$ required for an order-disorder transition increases by increasing the He concentration. Also note that $\mathrm{T_c}$--beyond which disordering was not observed--is $\sim 740$ at all He concentrations. At T$>$T$\mathrm{_c}$, mono-vacancies are mobile and contribute to re-ordering. The re-ordering rate increases exponentially, as the equilibrium mono-vacancy concentration dominates the chemically-induced vacancy concentration. Adding He has a very limited influence on this critical temperature. 

Another interesting feature of these maps is the independence of the order parameter with respect to the ordering temperature.  As shown in \ref{Fig:Experiment} (a), at T$<$T$\mathrm{_c}$ along the vertical dash line, the order parameter is constant. To explain this behavior, consider an ordered structure with He-V bubbles embedded in the system. Irradiation dissolves and scatters He atoms and vacancies in the system. Considering the formation energy, He-to-vacancy ratio and binding energy of different He-V complexes, stable and highly mobile $\mathrm{V_2He}$ complexes form immediately in the disordered volume. The diffusion of $\mathrm{V_2He}$ through the disordered lattice reorders the system. However, post-cascade clustering, a thermally activated phenomenon known as secondary bubble nucleation \cite{trinkaus2003helium}, decreases the number of  $\mathrm{V_2He}$ complexes by trapping them. Both diffusion and clustering are thermally activated phenomena for which the former results in re-ordering and the latter removes the $\mathrm{V_2He}$ complexes from the system. Therefore, the effectiveness of $\mathrm{V_2He}$ in re-ordering depends on the species' diffusion distance before sinking at the V-He cluster. This travel distance is independent of temperature and only depends on the He content. Therefore, the model is in agreement with the experimental measurements, and predicts a weak temperature-dependence of $ID_c$ in the 400K-740K range. 

Finally, the model predicts that at T$<$350 K, $\mathrm{V_2He}$ is not sufficiently mobile to recover from an irradiation dose greater than 0.1 dpa and a step should be observed in the order parameter contours. This limit is independent of He content; increasing the amount of He would not enhance the re-ordering.

Our findings show that adding several hundred appm of He postpones the order-disorder transformation, by enhancing re-ordering, a thermally activated process, which depends on the stability and mobility of ordering agents--in this case $\mathrm{V_2He}$. Adding 400 ppm of He increases $ID_c$ from 0.05 dpa to 2 dpa. Immiscible atoms stabilize non-equilibrium vacancies and reduce the re-ordering temperature significantly. This is a factor that should be considered in designing alloys when targeting properties arising from the stability of ordered structures, specifically in driven processes such as severe deformation and mechanical alloying. Furthermore, this mechanism may explain phenomena such as ordering of intermetallic ferromagnetic thin films of FePd under He ion irradiation below 600 K \cite{bernas2003ordering}, or enhancement of ordering by adding immiscible Ag to FePt \cite{takahashi2005low}.

 \appendix

 \begin{acknowledgments}
The authors thank Professor Jeffrey J. Hoyt from McMaster University and Dr. Enrique Martinez from Los Alamos National Laboratories for insightful discussion. Financial support for this work came from NSERC, a NSERC-UNENE Collaborative Research and Development (CRD) project and the NSERC/UNENE industrial Research Chair in Nuclear Materials at Queen's University. We thank Compute Canada for generous allocation of computer resources.
 \end{acknowledgments}

\bibliography{Biblio_Ordering.bib}

\end{document}